\documentclass[sigconf]{acmart}

\setcopyright{acmcopyright}
\copyrightyear{2023}
\acmYear{2023}
\acmDOI{XXXXXXX.XXXXXXX}

\usepackage{multirow}

\usepackage{xcolor}
\newif\ifcomment
\commenttrue
\ifcomment
\newcommand{\shirin}[1]{{\bf \textcolor{purple}{Shirin: #1}}}
\newcommand{\sayak}[1]{{\bf \textcolor{red}{Sayak: #1}}}
\newcommand{\ana}[1]{{\bf \textcolor{blue}{Ana: #1}}}
\else
\newcommand{\shirin}[1]{}
\newcommand{\ana}[1]{}
\newcommand{\sayak}[1]{}
\fi

\usepackage{caption}
\usepackage{subcaption}

\acmConference[The Web Conference (WWW) '23]{The Web Conference(WWW) '23}{April 30--May 4, 2023}{Austin, Texas}

\begin{document}

\title{Twitter Users' Behavioral Response to
Toxic Replies}

\author{Ana Aleksandric}
\affiliation{%
  \institution{University of Texas at Arlington}
  \city{Arlington, TX}
  \country{United States}}

\author{Sayak Saha Roy}
\affiliation{%
  \institution{University of Texas at Arlington}
  \city{Arlington, TX}
  \country{United States}}

\author{Shirin Nilizadeh}
\affiliation{%
  \institution{University of Texas at Arlington}
  \city{Arlington, TX}
  \country{United States}}


\begin{abstract}
Online toxic attacks, such as harassment, trolling, and hate speech have been linked to an increase in offline violence and negative psychological effects on victims. In this paper, we studied the impact of toxicity on users' online behavior. We collected a sample of 79.8k Twitter conversations. Then, through a longitudinal study, for nine weeks, we tracked and compared the behavioral reactions of authors, who were toxicity victims, with those who were not. We found that toxicity victims show a combination of the following behavioral reactions: \emph{avoidance}, \emph{revenge}, \emph{countermeasures}, and \emph{negotiation}. 
We performed statistical tests to understand the significance of the contribution of toxic replies toward user behaviors while considering confounding factors, such as the structure of conversations and the user accounts' visibility, identifiability, and activity level. Interestingly, we found that compared to other random authors, victims are more likely to engage in conversations, reply in a toxic way, and unfollow toxicity instigators. Even if the toxicity is directed at other participants, the root authors are more likely to engage in the conversations and reply in a toxic way. 
However, victims who have verified accounts are less likely to participate in conversations or respond by posting toxic comments. 
In addition, replies are more likely to be removed in conversations with a larger percentage of toxic nested replies and toxic replies directed at other users. 
Our results can assist further studies in developing more effective detection and intervention methods for reducing the negative consequences of toxicity on social media. 
\end{abstract}





\maketitle

\section{Introduction}
\label{intro}
These days, social media is rampant with toxic attacks such as offensive language, trolling, and using hate speech. The primary goal of these attacks is to silence, insult or demoralize people, especially those belonging to already marginalized groups~\cite{tahmasbi2021go,fredericks2021waiting,matamoros2017platformed,AKHTAR2019322}. 
These attacks are usually targeted, e.g., as part of a smear campaign to damage or call into question someone's reputation~\cite{zannettou2019disinformation,ratkiewicz2011truthy,hannan2018trolling,im2020still,mihaylov2015finding}.
They can also be coordinated using other communication mediums and implemented by many users~\cite{zannettou2020measuring,tahmasbi2021go,doerfler2021m,chatzakou2019detecting}. 

Psychological research has studied the negative effects of online harassment, cyberbullying, and trolling on individuals' psychological states and well-being~\cite{hinduja2007offline, kowalski2014bullying, giumetti2013rude}, showing that they often cause overwhelming and stressful situations for the victims~\cite{doi:10.1080/17405629.2011.643170,AKHTAR2019322,arafa2017pattern,lenhart2016online,ojanen2014investigating}. These studies found that victims are more prone to show self-harming behaviors as well as suffer from depression and anxiety~\cite{EYUBOGLU2021113730,ijerph17010045,Alhajji}. To counter online toxic attacks, many social media platforms have started implementing content moderation mechanisms~\cite{Twitter2020rules, Facebook2020rules} that block accounts~\cite{geiger2016bot,carlson2020report} and remove content~\cite{chaudhary2021countering}. However, it is debatable whether they provide sufficient mitigation to the potential psychological damage caused to the victim as a result of the online toxicity~\cite{jhaver2018online,gerrard2018beyond}.

To the best of our knowledge, no work has conducted a longitudinal data-driven study to examine the impact of toxic content on victims' online behavior. In this paper, we study how the victims \textit{respond} to toxic content in terms of their behavioral actions on Twitter. Our goal is to identify key factors which accurately represent the scale at which toxic replies impact victims. 
A recent study on online cyberbullying~\cite{ericsti2019reactions} targeting undergraduate and high school students found that victims primarily engaged in four types of behavioral reactions towards such attacks: \emph{avoidance}, \emph{revenge}, employ \emph{countermeasures}, and \emph{negotiation}. We use this as a framework for creating meaningful groupings of behavioral responses to toxic content. For example, we examine if the victims try to avoid further encounters with toxic content, e.g., by removing their posts or even deleting their accounts (which are also signs of being silenced), or if they tend to negotiate by posting comments in conversations, and even take revenge by responding in a toxic way, or employ countermeasures by ignoring such content but unfollowing the toxicity instigators. In our analysis, we consider factors that might have an impact on users' social media behavior, including the number of toxic replies in conversations, the structure of conversations, the location of the toxic content in the conversations tree, and the social relationship between conversations' participants (i.e., if they follow each other). We also explore the effects of account-specific attributes on users' decisions, including their online visibility, identifiability, and activity level. 

For our analysis, we collected a random sample of 79.8k Twitter conversations from August 14th to September 28th, 2021. We used this data to identify users involved in these conversations, identify conversations with toxic replies, and finally collect longitudinal data on users' online behavior. 
We represented the structure of a conversation using a reply tree, which encodes the relationships between tweets, where two tweets are connected if one is a reply to the other. We used Google Perspective~\cite{jigsaw2021perspective}, a natural language-based AI tool used to identify toxicity in text, to detect conversations that received toxic replies. 
Finally, we analyzed and characterized the behavioral reactions of root authors receiving toxic replies, who we call \emph{toxicity victims}, and compared their behavior with those of root authors receiving \emph{no} toxic replies, who we call \emph{random authors} and are our control group.  
We formulated the following hypotheses and performed appropriate statistical tests to understand the significance of the contribution of toxic replies toward user behaviors:
\textbf{H1}: Toxicity victims are more likely to deactivate their accounts compared to random authors. 
\textbf{H2}: Toxicity victims are more likely to switch their accounts to private mode. 
\textbf{H3}: Toxicity victims are more likely to engage in conversations.
\textbf{H4}: Toxicity victims are more likely to engage in conversations if receiving toxic replies from a larger number of toxicity instigators.
\textbf{H5}: Toxicity victims are more likely to respond back in a toxic way. 
\textbf{H6}: Toxicity victims are more likely to respond back in a toxic way if receiving toxic replies from a larger number of toxicity instigators. 
\textbf{H7}: Toxicity victims are more likely to delete their original posts compared to random authors.
\textbf{H8}: Replies in conversations with toxic replies are more likely to be deleted compared to conversations without toxic replies.
\textbf{H9}: Toxicity victims are more likely to unfollow toxicity instigators compared to random authors. 

Our study yields multiple important findings. 
We observed that different users respond differently to toxic content and identified groups of users who show similar reaction patterns. For example, in terms of \emph{disregard and avoidance}, we demonstrate that 30.96\% of victims ignored toxic content and did not show any reactions.  
In terms of \emph{negotiation}, we found toxicity victims compared to others are more likely to engage in the conversations (60.3\% vs. 47.4\%), and 12.4\% of victims employed \emph{countermeasures} by unfollowing toxicity instigators. Analyzing the contributing factors to users' behavior, our results suggest that users who receive toxic \emph{direct} replies are less likely to engage in the conversation, while users who receive \emph{nested} toxic replies engage more in conversations with toxic replies and tend to have toxic responses. 
These show that the location of toxic content in the conversation can affect victim's reactions. 
Interestingly, \emph{verified} accounts are less likely to engage in conversations with toxic replies, indicating that the social status of users can have an impact on how they perceive toxic content and react to them. Finally, identifiable accounts are more likely to engage in conversations by responding in a toxic way.
\begin{figure*}[t!]
\centerline{\includegraphics[width=0.85\textwidth]{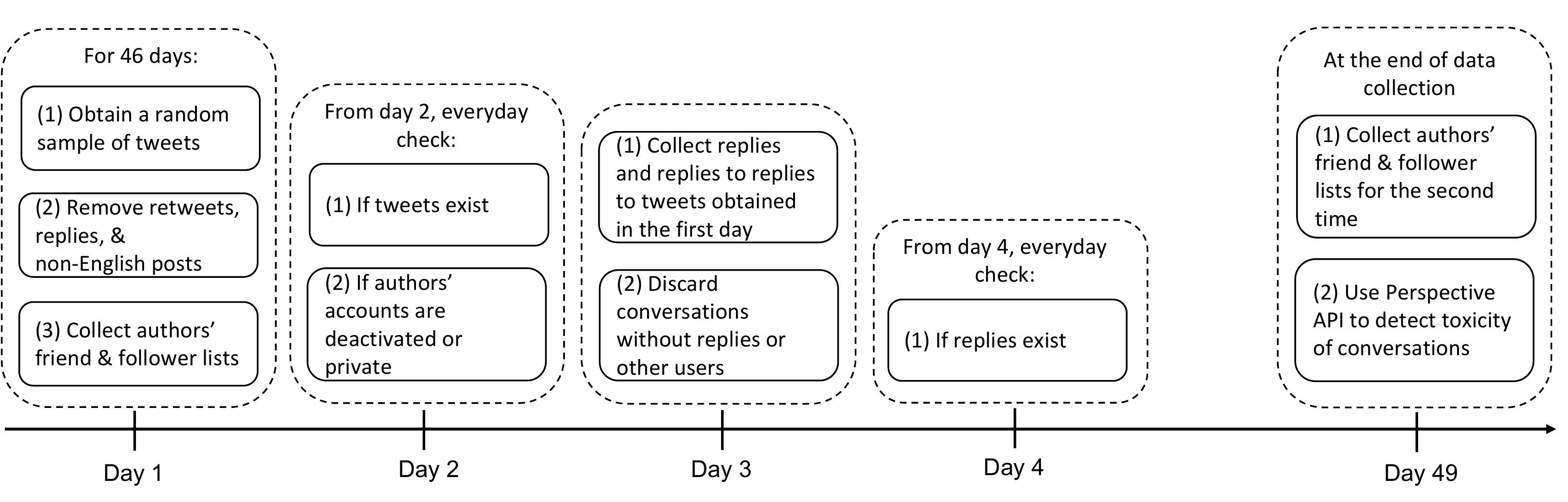}}
\caption{Data collection process}
\label{data-collection}
\end{figure*}

\section{Related Work}
Generalized online abuse encompasses several unhealthy online behaviors, including attacks of racism against minority communities~\cite{tahmasbi2021go,fredericks2021waiting,matamoros2017platformed}, misogynistic hatred~\cite{mantilla2013gendertrolling,parent2019social} and toxic masculinity~\cite{southern2019othering}, which are aimed at groups of vulnerable individuals. 
In recent years, online abuse have received a lot of attention. Here, we focus on studies that characterize online abuse and investigate its psychological impact of them on humans. To the best of our knowledge, there is no work that has studied the consequences of receiving toxic replies on users' online behavior. 

\textbf{Prevalence of Social Media Victimization.} 
Since abuse and toxicity are often carried out with the intention to humiliate or manipulate targeted individuals~\cite{varjas2010high,chraska2012victim,parent2019social,erjavec2012you}, social media is often considered to be the chief outlet for housing such attacks due to high visibility~\cite{yang2014social,yang2021social,duffy2019gendered} and more opportunities to remain anonymous~\cite{schlesinger2017situated,correa2015many,zhang2014anonymity}. 
Online abuse and toxicity can be influenced by several socioeconomic factors. 
Prior work has examined the association between abuse and on-the-ground ``trigger'' events, \textit{e.g.,} terrorist attacks, 
and political events~\cite{williams2015cyberhate, olteanu2018effect,hine2017kek, harlow2015story}.
The prevalence and characteristics of hate speech have also been studied on specific web communities, such as r/Gab~\cite{zannettou2018origins, zannettou2018gab}, 4chan's Politically Incorrect board (/pol/)~\cite{hine2016kek}, Twitter~\cite{elsherief2018hate,chatzakou2017mean, zannettou2020quantitative}, and Whisper~\cite{silva2016analyzing}. Some works have shown that online abuse and toxic comments are  normalized in several communities~\cite{beres2021don, mantilla2013gendertrolling, parent2019social}. 
There have also been a few efforts to understand the characteristic differences between hate targets and hate instigators~\cite{elsherief2018peer,ribeiro2018characterizing,elsherief2018hate,10.1145/3485447.3512131}. 

\textbf{The Psychological Impacts of Online Abuse.} 
Established literature on the psychological impacts of online abuse is mostly focused on cyberbullying. A study determined that middle and high school students who had been cyber-bullied were more prone to exhibit self-harming behavior as well as suffer from depression, anxiety, and lower self esteem~\cite{EYUBOGLU2021113730}. Similar studies~\cite{jenaro2018systematic,brack2014cyberbullying,MARTINEZMONTEAGUDO2020112856,Alhajji} found the same patterns of psychological distress in young adolescents~\cite{ijerph17010045} and adults~\cite{wang2019common}, due to cyberbullying attacks. Victims of online harassment might also respond by acceptance and self-blame~\cite{doi:10.1177/1461444818781324}, with those having lower psychological endurance being more vulnerable to emotional outbursts. Some studies explored the effects of trolling on victims~\cite{AKHTAR2019322}, and identified the circumstances when victims are more likely to respond to trolls~\cite{SUN2021106786}.

\section{Data Collection} 
Figure~\ref{data-collection} shows the pipeline used for collecting and processing our datasets. Broadly, we aimed to (a)~obtain a random set of conversations, and (b)~track the twitter activities of root authors. 

\textbf{Daily Collection of a Random Sample of Twitter Conversations:} 
We used the Twitter API~\cite{twitterapi} to collect a 1\% random sample of tweets for 46 consecutive days, starting from August 14th till September 28th, 2021, and extracted English tweets which are not retweets or replies belonging to other conversations. 
For each initial tweet, we waited at least two days before collecting the replies of the conversation. For example, if we collected a random sample of tweets on September 1st, we would start collecting replies for each of these tweets on September 3rd. This is to give enough time so that the initial tweet can turn into a conversation. However, we discarded many tweets that did not receive any comment or had been deleted by the time we attempted to collect their replies. 
We also removed the conversations where the replies for the tweets were all posted by the author of the initial tweet. 

\begin{figure}[t]
\centerline{\includegraphics[width=0.6\columnwidth]{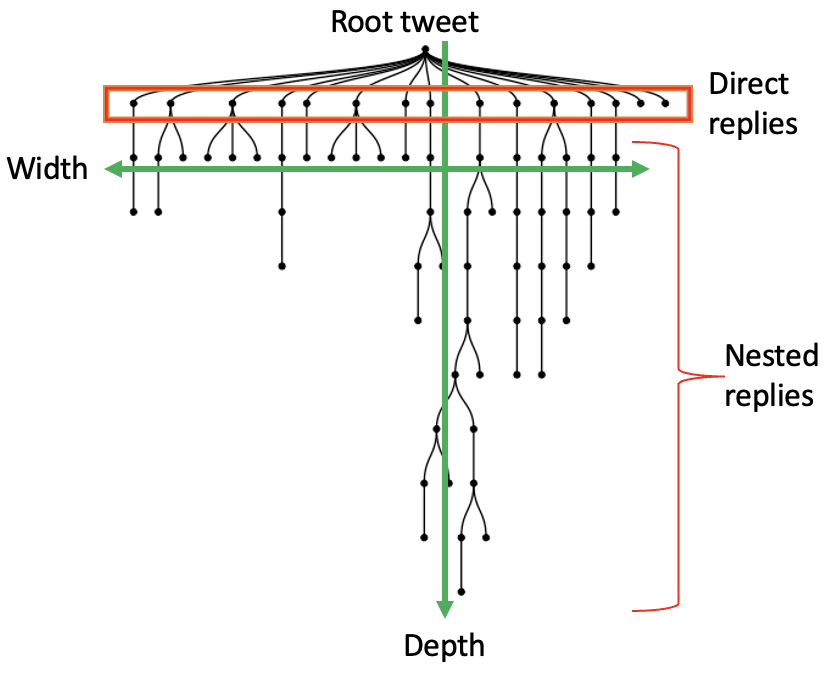}}
\caption{An example of conversation tree.}
\label{tree2}
\end{figure}

\textbf{Conversations, Reply Trees and Some Definitions:} We used Twarc~\cite{twarc}, a Python wrapper for the official Twitter API to obtain the entire conversation for each initial tweet, including its direct replies and nested replies (replies to replies). 
As it is shown in Figure~\ref{tree2}, we represent each conversation as a \emph{reply tree}, where one tweet is \emph{child} of another tweet when one is a reply to the other. The initial tweets represent the roots of \emph{reply trees}. We call the author of the root tweet as  \emph{root author}. We also define \emph{direct replies} as the first level replies in a conversation, which is the set of replies to the root tweet. 
\emph{Nested replies} refer to other levels of replies in a conversation other than the first layer, i.e., replies to replies. 

As it is shown in Figure~\ref{tree2}, each conversation as a \emph{reply tree} has the following properties: \emph{Size}, which indicates the number of tweets in the conversation; \emph{Depth}, which is the depth of the conversation's deepest node; \emph{Width}, which is the maximum number of nodes at any depth in the tree. 

\textbf{Pre-processing and Filtering:} Getting the replies for tweets of each day can take a long time, up to couple of hours. Therefore, the first tweets collected for the certain day would have much less time to receive any comments than last tweets obtained from that day. In order to solve this problem, we only kept replies sent in the first 48 hours after the root tweet was posted. 
Also, some tweets contained only links, images, and videos instead of text.
Since our approach of detecting toxic tweets is text-based, such conversations were removed from the dataset. 
Moreover, certain \emph{reply trees} were missing some replies due to the errors received during their data collection. Since these errors can have multiple reasons including, replies being deleted or hidden by their authors, or root authors, we removed trees containing such errors from our dataset. 
Furthermore, we noticed that not all the root authors in our sample are unique, therefore to avoid duplication in our statistical analysis, we randomly selected a single conversation from each root author. 
\textbf{Our Conversation Dataset:} Finally, our dataset consists of 79,799 conversations with 528,041 tweets, posted by 328,390 unique users, out of which 79,799 are root authors.

\textbf{Tracking activity of root authors:}
For each root author and root tweet, we kept track of the following activities:
\emph{account deactivation}, \emph{account privatization}, and \emph{tweet deletion}, as well as root authors' \emph{lists of followers and friends}. In detail, as it is shown in Figure~\ref{data-collection}, from the second day of our data collection, we checked on \emph{a daily basis} if the root tweets and authors still exist using Twitter API. Similarly, from the fourth day of our data collection, after collecting the replies to a root tweet, every day we were checking if the replies still exist. When a certain tweet or account is not accessible, API returns helpful error codes, e.g., \emph{Sorry, you are not authorized to see this status}, \emph{User not found}, and \emph{No status found with that ID}~\cite{twitter-errors}. These error codes indicate that the user account has become private, been deactivated and the tweet has been deleted, respectively. Note that we cannot certainly identify who deleted a particular tweet. For instance, the root tweet can be deleted by the root author, or in case the deleted tweet is a comment, it can be deleted by the user who posted that specific comment or by the root author. 
Even though this process was repeated daily during the entire process of data collection, the data collection failed on some days. 
More precisely, in the analysis, we focus on the presence of the tweets and accounts three days after the corresponding root tweets were collected, but the data about their presence is missing in 7.6\% of the conversations, with very similar distributions in conversations with and without toxic replies. Therefore, in the  statistical models related to tweet deletion, account deactivation, and account privatization, we omit these conversations. 

As illustrated in Figure~\ref{data-collection}, we collected the followers and friends lists of all root authors every day, once the daily random sample is obtained. Thus, the lists collected at this time represent a snapshot of the authors' friends/followers lists before their tweets turned into a conversation. 
Furthermore, around 49 days after the first day when root tweets were collected, we again collected the list of friends and followers for all root authors. 
This is to analyze the impact of toxicity on users' online relationships. Also, note that in the analysis we report the distribution of the percent of toxicity instigators that the victims unfollowed. As toxicity instigators do not exist in other conversations, comparing the unfollowing ratio between victims and other random users was not applicable. 
We could not collect the relationships, i.e., follower and friends' lists, right after obtaining the conversations, because of the number of API calls we could issue every day. 
This delay in collecting relationships imposes some limitations as users might end followership and friendships due to other events during this time. However, in our analysis, we compare these variables of toxicity victims and random authors and seeing a difference can be an indicator of the impact of toxicity on Twitter relationships. 
From our \emph{unfriend} analysis, we had to discard 2,488 conversations, where 403 and 2,085 are conversations with and without toxic replies respectively because we were not able to obtain the friends' list for their root authors due to the authors either making their accounts private or deactivating them. 

\textbf{Identifying Conversations with Toxic Replies:} We used Google's Perspective API~\cite{jigsaw2021perspective} to detect toxic replies in our dataset. 
This AI-based tool investigates whether a provided text contains language indicating abusive or inappropriate attributes such as \emph{Severe toxicity}, \emph{Profanity}, \emph{Sexually expletives}, \textit{Threats}, \textit{Insults} and assigns a score between 0 to 1, with a higher score indicating more severity for a particular attribute. In this paper, we are only considering scores for the \emph{Severe Toxicity} attribute since Google's Perspective API defines a text having this attribute to as rude, disrespectful, or unreasonable comment which meets the general standards for a comment which might be considered to be hateful or toxic. 
In our analysis, for finding the number of toxic comments, we needed to create a binary variable that labels a tweet as toxic or not toxic. We labeled a tweet as toxic, if its severe toxicity score is higher than 0.5. To determine this score threshold, two independent coders labelled 200 random tweets with an inter-rater agreement of 94\% and a Cohen's kappa of 0.69. The inter-rater reliability showed substantial agreement for \textit{severe toxicity} between the manual coders and the Perspective API for scores around 0.5.

From 528,041 tweets in our dataset, 34,208 (6.5\%) tweets were labelled as \textit{toxic}, where about 52\% of toxic tweets were directed to the root authors, 38.4\% of toxic tweets were written by the root authors. Only 9.6\% of toxic tweets were posted by other users and directed to other conversation participants. From all conversations, 10,953 conversations (13.7\%) have received toxic replies posted by other users and directed to the root author. We call such conversations as \emph{Conversations of Toxicity Victims (CTV)}, and the rest of conversations is named \emph{Conversations of Other Authors (COA)}, which might include toxic replies that are written by the root author or are directed to others. However, our focus is to investigate the way root authors respond to toxic replies that are directed to them only.

\section{Variables and Statistical Models} 
We use multivariate regression to study the associations between toxicity and Twitter users' behaviors. To provide a larger context for interpretation of our analyses, we compare the users' behaviors of conversations \emph{with} and \emph{without} toxic replies.  
Below, we explain our dependent, independent, and control variables in detail.

\subsection{Dependent Variables} \label{dep_vars}
These variables represent our understudy user behaviors, including: 
(1)~\emph{deactivated}: a binary variable indicating whether the account is deactivated. Account deactivation can be considered an extreme reaction to toxicity, when someone decides to leave the platform temporarily or permanently, and can be considered as an \emph{avoidance} behavior. 
Although we collected the data about presence of the accounts for nine weeks, we decided to use the data obtained on the third day after collection of the root tweets, because most of the root authors deactivated their accounts on that date (See Figure~\ref{daily_deact} in Appendix~\ref{appendixA}). 
(2)~\emph{private}: a binary variable showing whether the account is gone to the private mode (i.e., only the root author's followers can observe and interact with their tweets).  
This can be considered an severe reaction to toxicity, when someone decides to make their account private and limit their visibility. Similarly, this can be considered as an \emph{avoidance} behavior. 
We assigned values to this variable using the data from day three for the analysis, because most of the accounts went to the private mode on that date (See Figure~\ref{daily_private} in Appendix~\ref{appendixA}). 
(3)~\emph{root\_tweet\_deleted}: a binary variable indicating whether the root tweet is deleted. Some users might delete their initial posts once they receive toxic replies to \emph{avoid} further toxicity. To assign values to this variable, we used the data collected three days after the collection of root tweets, because most of the root tweets were deleted on that date (See Figure~\ref{ddaily_deletion} in Appendix~\ref{appendixA}). 
(4)~\emph{\#deleted\_replies}: a numeric variable showing the number of deleted replies. 
Some users might delete the replies within the conversation to \emph{avoid} the situation. The information about replies' presence used in the analysis was based on the data collected three days after the collection of initial tweets. 
(5)~\emph{\#root\_author\_replies}: a numeric variable indicating the number of root authors' replies. In contrary to deleting tweets and replies, some users might decide to \emph{negotiate} about their point of view and participate in the conversation. This variable can help us capture this behavior. 
(6)~\emph{\#root\_author\_toxic\_replies}: the number of root authors' toxic replies per conversation. This can capture the \emph{revenge} behavior, when the victim responses to the toxicity in a toxic way. 
(7)~\emph{unfollowing}: a binary variable that shows if the user unfollows another user. This can be considered as either \emph{avoidance} or \emph{revenge} behavior, as the victim tries to avoid further communication with the toxicity instigator. 

\subsection{Independent Variables} 
Our independent variables are computed based on toxicity scores obtained for replies in the conversations. 
We could define the \emph{overall toxicity} as the ratio of number of toxic replies to the size of the conversation. 
However, not all toxic replies are directed at the root author and sometimes they are directed at other users posting in the conversations. 
Also, the location of toxic replies in the tree structure can affect the victims' reactions. For example, as it is shown in Figure~\ref{tree2}, direct replies might be more visible compared to nested replies.   
Therefore, we considered three independent variables considering such factors: 
(1)~\emph{direct\_toxicity}, a numeric variable defined as the ratio of the toxic direct replies, targeting the root author, to the total number of direct replies.  
(2)~\emph{nested\_toxicity}, a numeric variable defined as the ratio of the number of toxic nested replies, targeting the root author, to the total number of nested replies in the conversation. 
(3)~\emph{toxicity\_to\_others}, a numeric variable defined as the ratio of the toxic replies, directed at the other conversation participants, to the total number of toxic replies in a conversation. 

\subsection{Control Variables}
When comparing the behavior of root authors in conversations \emph{with} and \emph{without} toxic replies, i.e., \emph{victims} and \emph{other authors}, we controlled for features that might have influenced their behavior in conversations (i.e., confounding factors). 
In particular, we controlled for features that capture the structure of conversations, the relationship between root author and toxicity instigator, and users' \emph{activity}, \emph{visibility} and \emph{identifiability} on Twitter.    

\textbf{Online activity} includes \emph{num\_friends}, \emph{num\_tweets} and \emph{account\_age} (in years) as numeric variables.
If a user posts many tweets then they might also tend to engage in conversations more, even if they receive toxic replies, or if a user is more established, i.e., been using Twitter for more years, then they might perceive toxicity and react to it differently. 
\textbf{Online visibility} includes  \emph{num\_followers}, and \emph{listed\_counts} as numeric variables and \emph{verified} as a binary variable.  
While other works~\cite{elsherief2018peer} have shown that online visibility and receiving hate are related, they might also influence users' reactions towards toxicity. For example, verified users or a user with more followers might get less negatively affected by receiving a few toxic replies, compared to a user who is not verified or has a few followers. 
\textbf{Identifiability} represent features from the user profiles which can help identify a user, including  \emph{description\_length} (in chars) as a numeric variable, and \emph{has\_URL}  and \emph{has\_location} as binary variables. We also captured \emph{has\_image} but noticed that all accounts are with profile images and therefore discarded this variable. 
We argue that it might be that more identifiable users are more likely to get negatively affected when they are under attack by others compare to anonymous users. Also, other works have shown that anonymous accounts tend to show abusive behavior more than others~\cite{schlesinger2017situated,correa2015many,zhang2014anonymity}. 
\textbf{Conversation structure}  includes \emph{size}, \emph{width}, and \emph{depth} as numeric variables. These features can play a role in how victims respond to toxicity, e.g., a toxic reply buried in a nested conversation might not have the same negativity level as having one in a short conversation. In our analysis, we also controlled for \textbf{root\_toxicity}, which indicates if the conversations starts with a toxic tweet. Such authors probably could expect receiving toxic replies and they might react differently. 
\section{Victim's Behavioral Responses} 
We provide the descriptive statistics about conversations, author accounts and victim's behavioral reactions. 

\begin{table}[t] \centering 
\caption{Characteristics of conversations in our dataset.}
\resizebox{\columnwidth}{!}{%
\begin{tabular}{lllll|llll}
\hline
\hline
& \multicolumn{4}{c}{CTV} & \multicolumn{4}{c}{COA}  \\
\hline
 Characteristics & Min    & Median & Mean     & Max  & Min    & Median & Mean     & Max \\
\hline
Size & 2 &  6 & 15.88 & 1842 & 2 & 3 &  5.14 & 906  \\
Width         & 1 & 2 & 9.25 & 1825  & 1 & 1 & 2.53 &  901 \\
Depth          & 2 &  3 & 4.42 & 198 & 2 & 2 & 3.11 & 98    \\
No. users   &  2 & 3 & 10.56 & 1818  & 2 & 2 &  3.48 & 889  \\
Direct Toxicity & 0  & 0.3 & 0.46 &  1 & 0 & 0 & 0 & 0  \\
Nested Toxicity & 0 & 0 & 0.06 & 1 &  0 & 0 & 0 & 0   \\
Toxicity to Others & 0 & 0 & 0.03 & 0.93  & 0  & 0 &  0.01 & 1  \\
No. Toxicity Inst. & 1 &  1 & 1.5 & 113 &   0 & 0 & 0 & 0  \\
\hline
 \# conversations   & \multicolumn{4}{c}{10,953}        & \multicolumn{4}{c}{68,846}  \\
\hline
\end{tabular}
\label{stats} 
}
\end{table}

\textbf{Conversations' Characteristics}: 
Table~\ref{stats} compares the characteristics of CTV and COA. Since the distribution of features is not normal, we present min, mean, median, and max values.
It shows that the prevalence of toxic direct replies to the root author ($Mean=0.06$) is higher compared to nested replies ($Mean=0.008,$). The Mean number of toxicity instigators is 0.21, while the Mean in CTV is 1.5, and the maximum number of instigators in our dataset is 113. 
The min, mean, median, and max values for conversation size in our dataset are 2, 3, 6.6, and 1,842, respectively. These results show that most conversations are small, however, the size of 8,872 (11.1\%), 291 (0.4\%), and 25 (0.03\%) of conversations is more than 10, 100, and 500.  
We also see that size, width and depth values are higher for CTV, e.g., the median and mean of size for CTV are about 6 and 16, while those for COA are about 3 and 5. 
The number of unique users who participated in the conversations is related to the conversation size because a conversation with more users has more posts. Similarly, we observe on average more users participate in CTV (Mean is about 11) compared to COA (Mean is about 3). 
We ran Mann-Whitney U tests for all the variables and found that there are significant differences between these characteristics in CTV and COA. For brevity, we do not present these results.

\begin{table}[t] \centering 
\caption{Comparing victims and other root authors' behaviors and their account characteristics.}
\label{reactions}
\resizebox{\columnwidth}{!}{%
\begin{tabular}{lllll|llll}
\hline
\hline
      & \multicolumn{4}{c}{Toxicity Victims} & \multicolumn{4}{c}{Other Root Users}   \\
\hline
   & Min    & Median  & Mean     & Max & Min    & Median  & Mean     & Max \\
\hline
Deactivated & 0 & 0 & 0.01 & 1 & 0 & 0 & 0.004 & 1  \\
Private & 0 & 0 & 0.006 & 1 & 0 & 0 & 0.008 & 1  \\
Author's Replies  & 0 & 1 &  2.13 & 313 & 0 &  0 & 1.009 & 504   \\
 Toxic Replies  & 0 & 0 & 0.21 & 23 & 0 & 0 & 0.05 & 26  \\
 Root Deleted & 0 & 0 & 0.018 & 1 & 0  & 0 &  0.024 & 1 \\
Del. Comments  & 0 &  0 & 0.0023 & 9 & 0 &  0 & 0.001  &  5 \\
 Unfollowed  & 0  &  0 & 0.13 &  1  & NA & NA & NA &  NA\\
\hline
\hline
No. followers       &  0  & 1,121  &  $\sim 70K$ 
&  $\sim 54M$ 
& 0 & 763  &  15427 & $\sim 36M$ 
\\
No. friends         & 0  & 571  &  1,857    & $\sim 1M$ 
& 0 & 533 & 1486 & $\sim 2M$ 
\\
No. tweets          &1  & 12679  & $\sim 36L$ 
& $\sim 1M$ 
& 1 & 10,595 &  31, 283 & $\sim 4M$ 
\\
 Verified                  & 0 & 0   & 0.07       & 1                & 0 & 0 & 0.038 & 1  \\
 Descr. len. (char) & 0 &  77  & 80.78 & 199      &  0 & 73 & 78.16 & 199 \\
 Listed count              & 0 & 7.0   & 298.1 & $\sim 210K$ 
 & 0 & 4 & 83.5  &  $\sim 117K$ 
 \\
 URL                       & 0 & 1 &  0.5 &  1  & 0 & 0 & 0.46 &  1  \\
 Location                  & 0 & 1 &  0.76 &   1          & 0 & 1 & 0.75 &  1 \\
 Acc. age (years)       &0 & 3 &  4.44  & 15        & 0 &  3 &  4.7 &  15 \\
\hline
 No. users     & \multicolumn{4}{l}{10,953}        & \multicolumn{4}{l}{68,846}       
\end{tabular}
}
\end{table}

\begin{figure}[t]
     \centering
     \begin{subfigure}{0.49\columnwidth}
         \includegraphics[width=\textwidth]{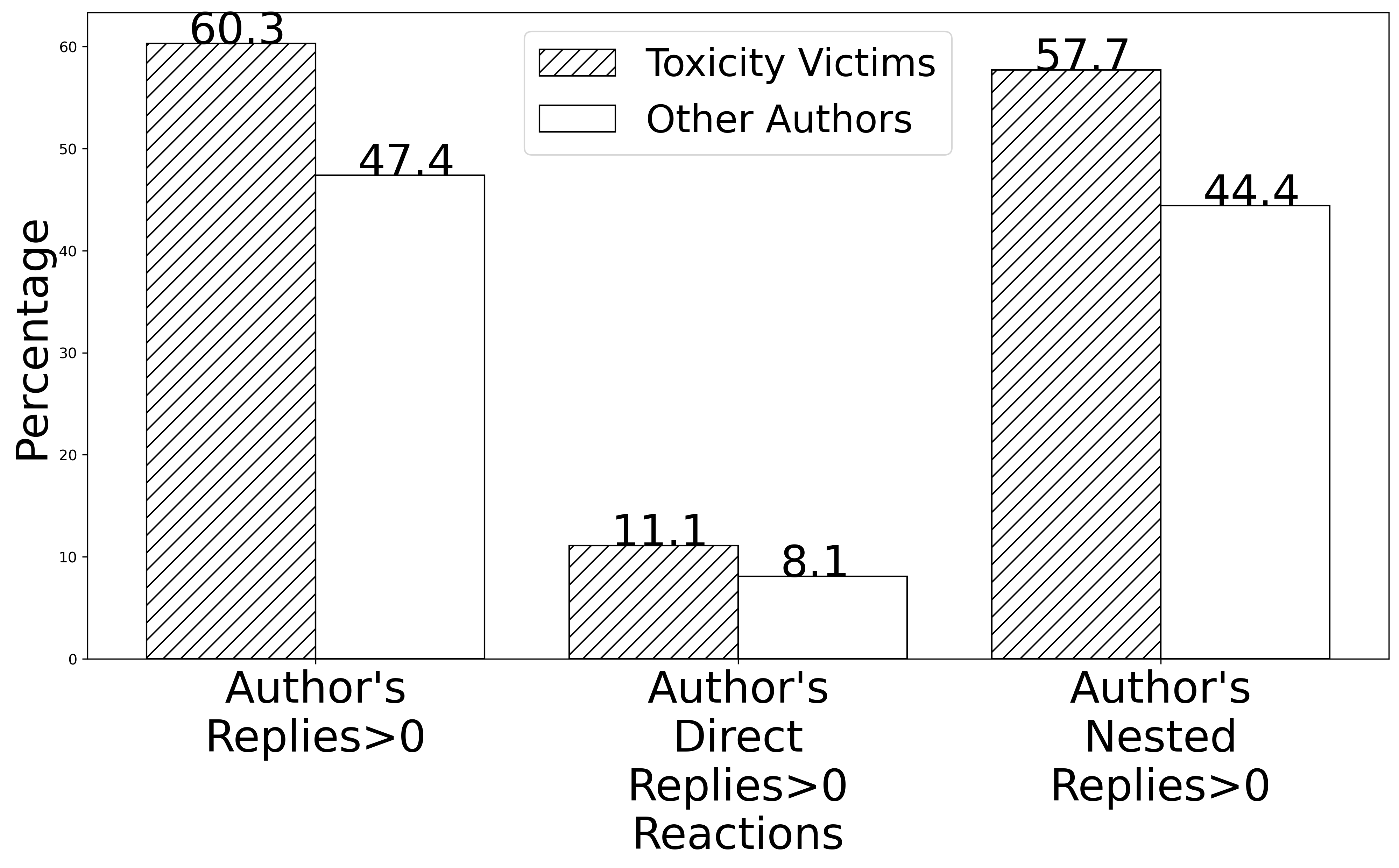}
         \caption{Root authors' engagement}
         \label{engagement}
     \end{subfigure}
     \begin{subfigure}{0.49\columnwidth}
         \includegraphics[width=\textwidth]{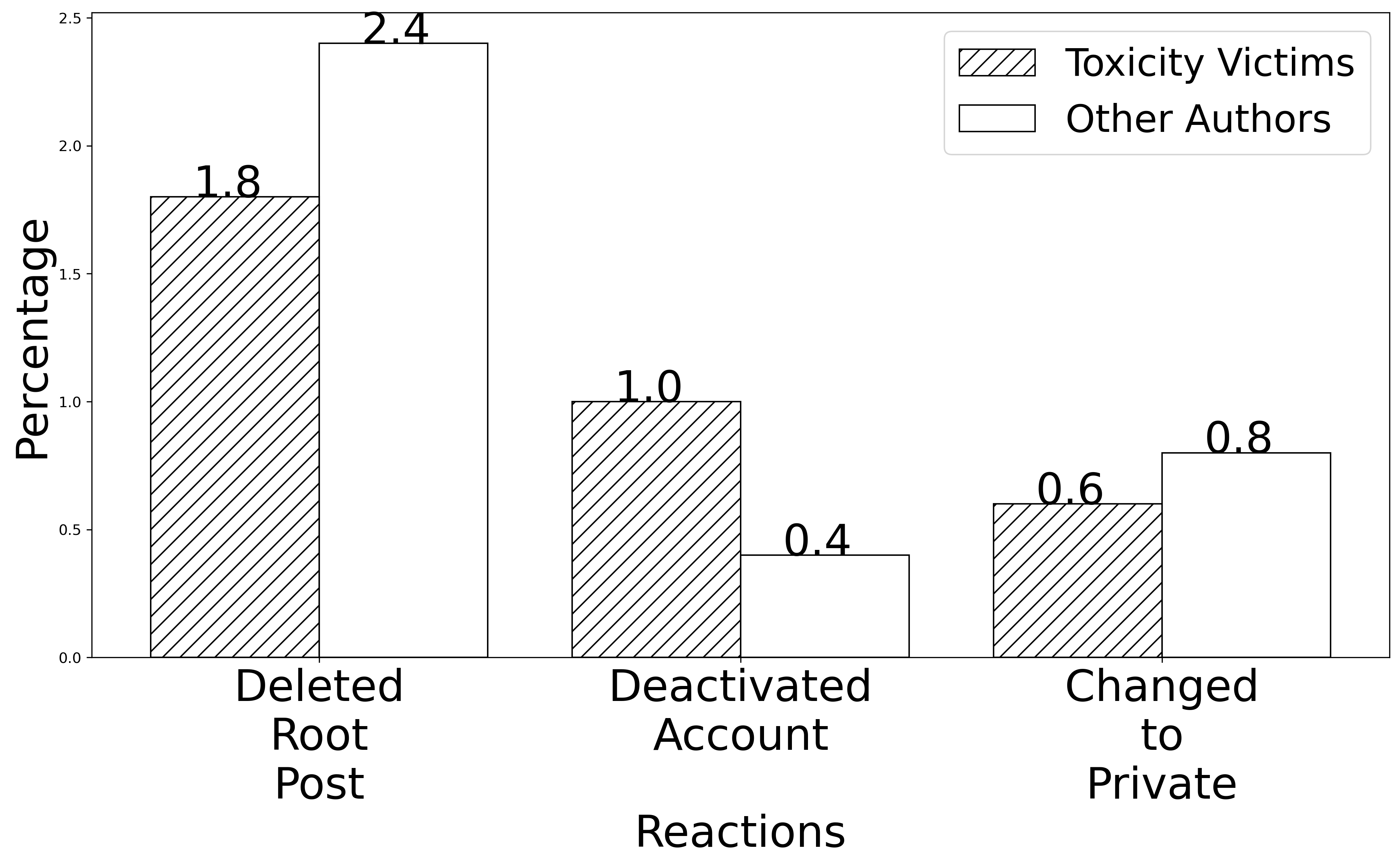}
         \caption{Authors' severe reactions}
         \label{deletion}
     \end{subfigure}

     \begin{subfigure}{0.49\columnwidth}
         \includegraphics[width=\textwidth]{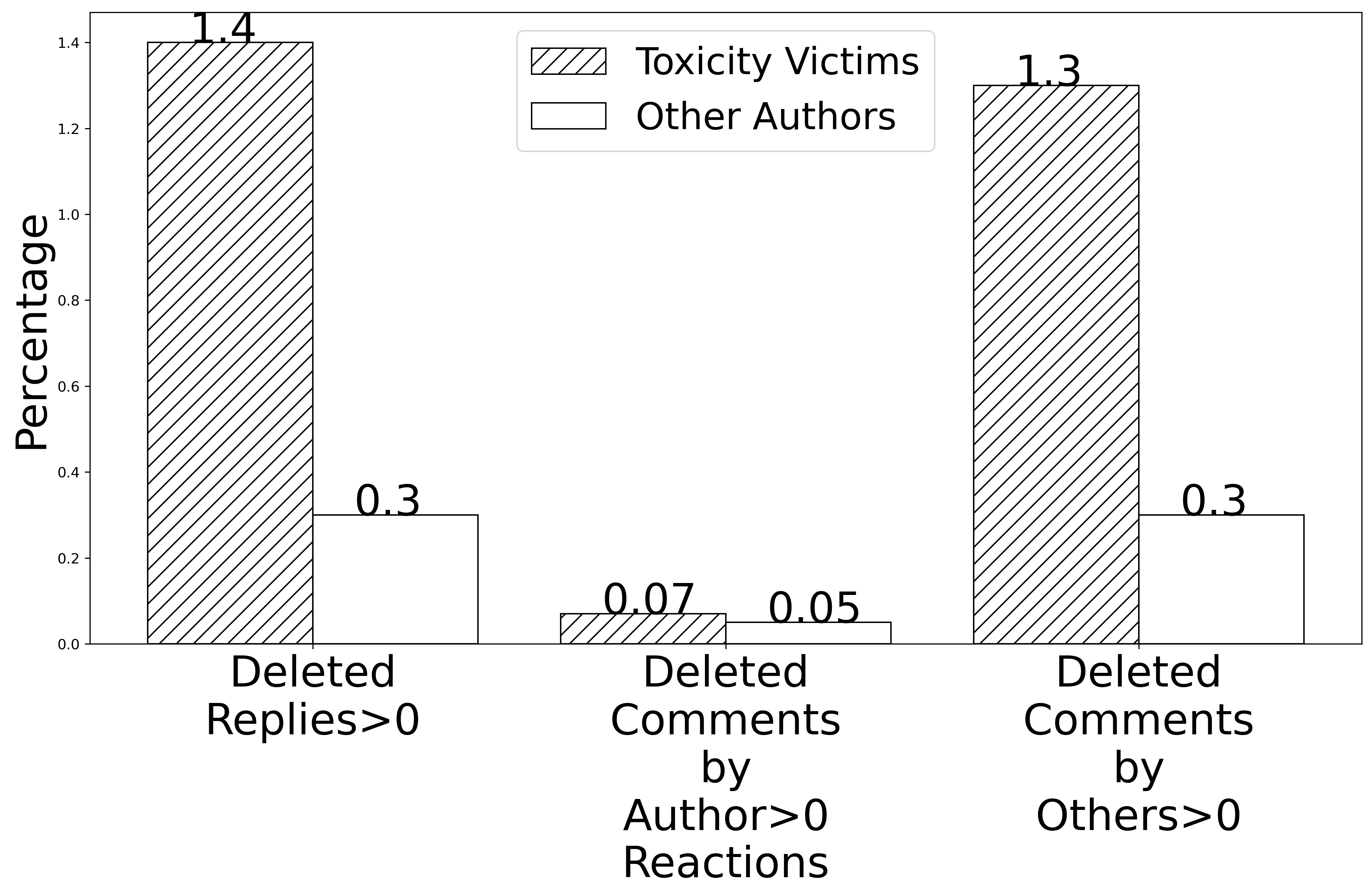}
         \caption{Deleted replies}
         \label{deleted_repl}
     \end{subfigure}
     \begin{subfigure}{0.49\columnwidth}
         \includegraphics[width=\textwidth]{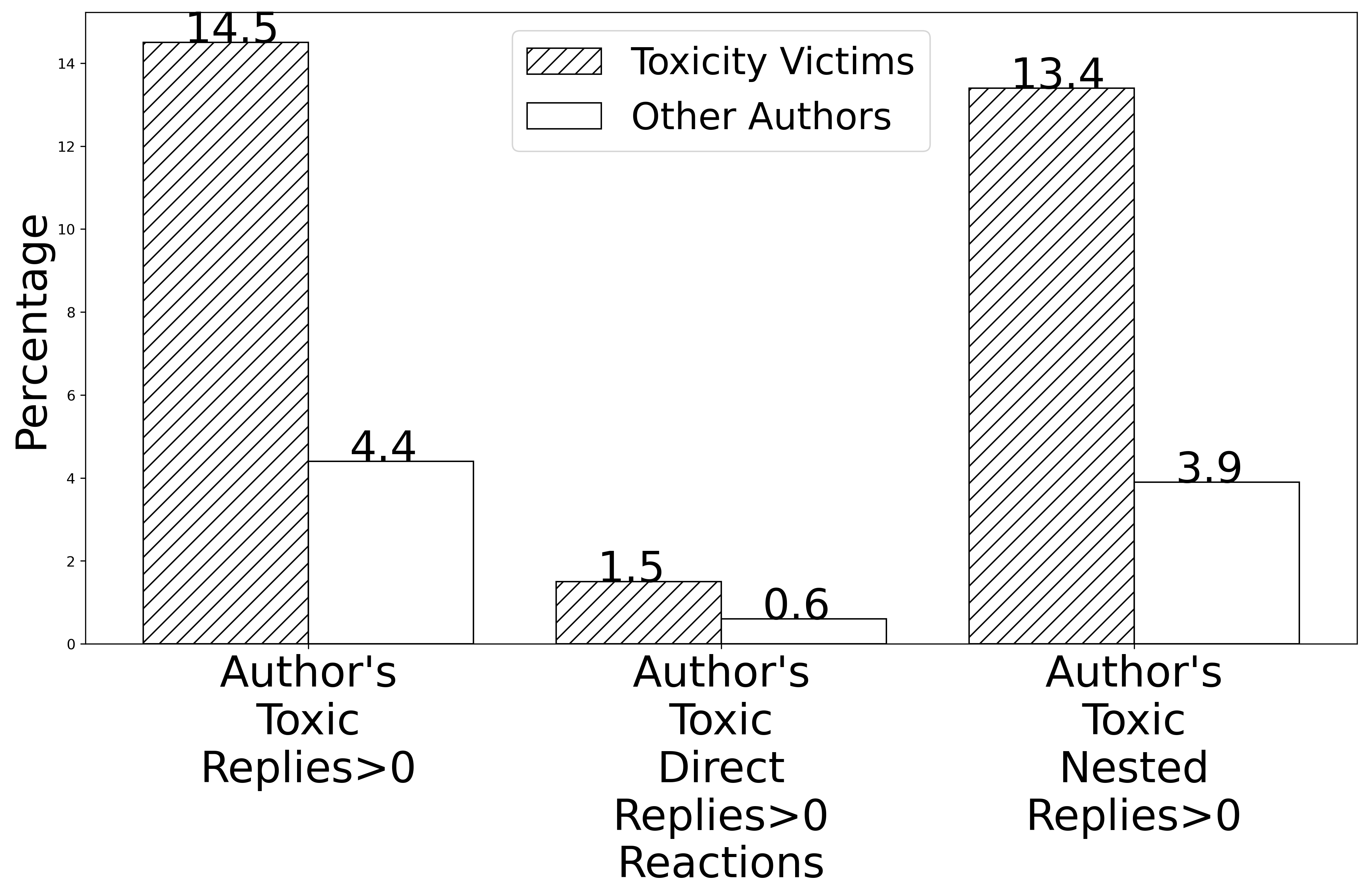}
         \caption{Root authors' toxic replies}
         \label{toxic-responses}
     \end{subfigure}
        \caption{Reactions among Toxicity Victims and other Users}
        \label{toxic_eng}
\end{figure}

\textbf{Victim's reactions:} Table~\ref{reactions} shows the statistics about the victim's reactions and identifies the more prevalent reactions. 
As illustrated in Figure~\ref{deletion}, percentage of victims who deactivated accounts or went private are 1\% and 0.6\%, respectively. 
Interestingly, Table \ref{reactions} shows that the maximum \emph{\#root\_author\_replies} is higher in COA compared to CTV (504 vs. 313), but the mean of \emph{\#root\_author\_replies} in CTV is approximately twice times the mean of \emph{\#root\_author\_replies} in COA (1.009 vs. 2.13). As shown in Figure~\ref{engagement}, toxicity victims are engaged in a larger percentage of conversations compared to other users (60.3\% vs. 47.4\%). Moreover, Figure~\ref{toxic-responses} reveals that the percent of CTV where the toxicity victims responded with at least one toxic reply is 14.5\%, while in COA, that number is 4.4\%. Furthermore, the percentage of all root authors' toxic replies that were immediate toxic replies to toxic comments is 3.1\% (the toxic children of the toxic node in the conversation tree). Finally, the percent of all children nodes where the root author directly responded to a toxic tweet which is toxic in nature is 13.4\%.

The percentage of victims that removed the root tweet is 1.8\% while the percentage of other users who deleted a root post is 2.4\% (Figure \ref{deletion}). Also, the mean of \emph{\#deleted\_replies} in CTV is 0.02, while in COA this number corresponds to 0.004, indicating that there might be a significant correlation between \emph{toxicity} of the conversation and number of deleted comments. 
The percentage of CTV where there is at least one removed comment belonging to the root author is 0.07\% while comment deletion occurred only in 0.05\% of COA (presented in Figure~\ref{deleted_repl}). 
We focused on deleted replies posted by the root author because authors can only delete their own tweets/replies. 
Interestingly, in 8 out of 137 CTV with at least one deleted reply, conversations contained at least one reply posted by the root author. In case of COA, in 35 out of 201 cases, conversations contained at least one removed reply belonging to the root author. This indicates that in most of the conversations, removed replies belong to other participants. 
Additionally, in our dataset, in 12.4\% of CTV, root authors unfollowed at least one user who posted toxic comments on their posts.
Finally, in 30.96\% of CTV, the root author did not respond in any way. 

\textbf{Root Authors' Account Characteristics:} 
Interestingly, Table~\ref{reactions} shows that our dataset contains a higher percentage of \emph{verified} accounts among the toxicity victims (0.07) compared to other users (0.038), consistent with the prior studies~\cite{elsherief2018peer}. In addition, the Mean values of followers, friends, tweets, and listed counts of toxicity victims are higher compared to these characteristics of other random root authors, again consistent with prior studies~\cite{elsherief2018peer}, indicating that toxicity victims might be more active and visible accounts compared to other users. Moreover, a higher percentage of victims have URL and location specified on their profiles, and the victims Mean account age might indicate that their accounts are younger compared to other users. We ran a Mann-Whitney test to compare the distributions of all account characteristics, and we found that there is a statistically significant difference between all account characteristics among victims and other users.

\begin{table}[t!] \centering
\caption{Descriptive Statistics of Cluster Characteristics.}
\resizebox{\columnwidth}{!}{%
\label{cluster-features}
\begin{tabular}{lp{0.03\textwidth}p{0.04\textwidth}p{0.05\textwidth}p{0.05\textwidth}p{0.05\textwidth}p{0.05\textwidth}p{0.05\textwidth}p{0.05\textwidth}}
\hline
No. & Size & \emph{Root deleted} & \emph{Switch to private} & \emph{Account deactivated} & \emph{Author replies} & \emph{Author toxic replies}  & \emph{Deleted comments} & \emph{Unfollow ratio} \\
\hline
1   & 1,042 &   0     &     0  &   0.012  &    0.21   &    0.003  &    0.002  &    0.96 \\
2    & 3,158 &    0    &   0 &   0.002 &      0.31   &    0.004  &    0.0004   &   0.002 \\
3   & 60 &   0   &    1 &   0.017   &     0.14  &     0.064  &     0  &    0.32 \\
4  &  234 &   0   &   0 &   0   &    0.37  &   0.48   &    0.002  &   0.95 \\
5   &  182 &    1    &      0 &   0.02 &       0.03  &    0.007   &     0.006  &   0.19 \\
6   &  1,072 &   0  &     0 &   0.007  &     0.35    &       0.48   &    0.0002  &      0.001 \\
7    & 4,337 & 0 &      0 & 0.0046 &   0.017         &       0.002    &  0.001  & 0.006 \\
\hline
\end{tabular}
}
\end{table}

\begin{figure}[h]
\centerline{\includegraphics[width=0.5\columnwidth]{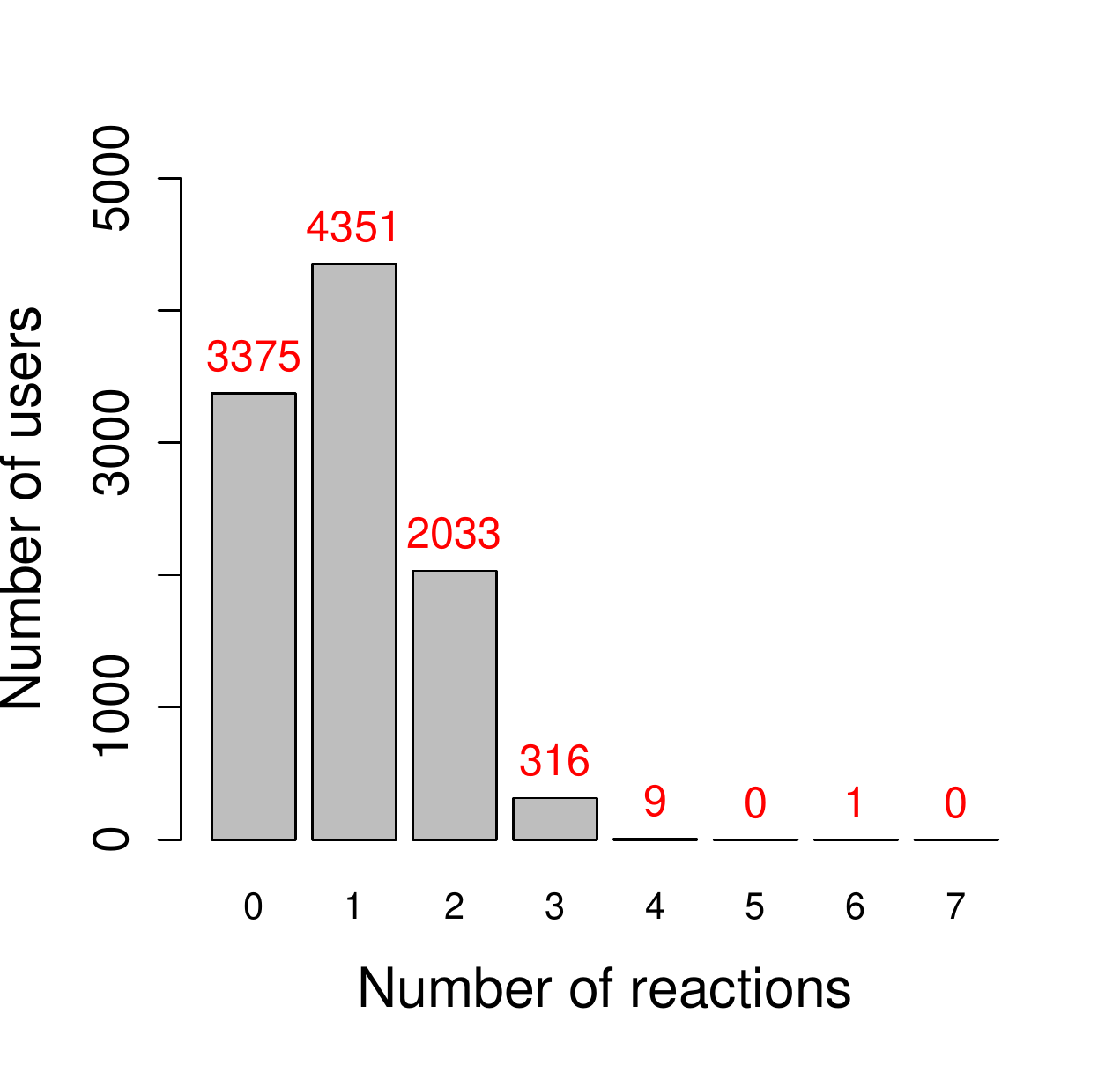}}
\caption{Histogram of victims' behavioral reactions.}
\label{num-reactions}
\end{figure}

\textbf{Clustering Victims by their Reactions}:   
Victims might show a combination of behavioral reactions, and such combinations might also be common among some groups of victims. Figure~\ref{num-reactions} shows the histogram of the number of behaviors a victim has shown in our dataset. The victim can have up to seven reactions, corresponding to the number of dependent variables described in Section~\ref{dep_vars}. 
As you can see, a huge number of victims (30.96\%) decided to ignore the toxic replies and did not respond, while 39.7\% of victims showed only one of our studied behavioral reactions. It is interesting that about 30\% of the victims showed a combination of these behaviors. In particular, 18.6\% and 2.9\% of victims showed a combination of two and three different behaviors, respectively. Only ten victims showed more than three different reactions in our dataset, specifically 9, 0, 1, and 0 showed four, five, six, and seven behaviors, respectively. 

We further employed k-means clustering to group victims using the seven behavioral reactions as features. While features such as \emph{deactivated}, \emph{private}, \emph{deleted} are binary variables, the rest are numeric variables. 
For this analysis, we discarded 868 (7.9\%) of CTVs, as for these conversations we failed to collect data for account and tweet existence three days after initial tweets were collected.

Since we could assign any $k$ for clustering, we examined clustering with various k values, starting from 4 clusters. We  increased the number of clusters by one, as long as we would see that the results create a meaningful cluster that is based on certain reactions. However, testing $k=8$ or more clusters produced very small clusters of victims but did not create a new grouping of reactions. Consequently, we set the number of clusters to \emph{seven}. 
As shown in Table~\ref{cluster-features}, each cluster is a combination of one or more users' behavioral reactions. The values in the table show the percentage of victims that have those reactions. 
Interestingly, 96\% of victims in cluster one ($N=1,042$) unfollowed toxicity instigators and engaged in the conversation positively because 21\% of victims in this cluster replied and only 0.3\% of them replied in a toxic way. Also, victims in cluster four ($N=234$) unfollowed toxicity instigators and engaged in the conversation \emph{negatively} because 37\% of victims in this cluster have replied and 48\% of those replied in a toxic way. Interestingly, users who are grouped in cluster seven were the ones who ignored toxic comments on their tweets, and this represents the largest cluster in the dataset ($N=4,337$). The second largest cluster, cluster two ($N=3,158$), represents victims who only engaged in conversations positively. The smallest cluster, cluster three ($N=60$), includes victims who all went private, about 14\% of them also engaged in conversations, and 32\% of them unfollowed toxicity instigators. Finally, all victims in cluster five have deleted the root tweet and 19\% of them unfollowed toxicity instigators, suggesting that these two behaviors might coexist. Cluster six includes 1,072 root authors who responded in a toxic way. 

In summary, this analysis confirms that victims can be grouped based on their behavioral reactions, and that some combinations of behaviors tend to appear together. For example, we observed that unfollowing is more likely to be used along with other reactions, as it was seen with user engagement (both positive and negative) in clusters 1 and 4, deleting in cluster 5, and going private in cluster 3.

\section{Hypotheses Testing}
We used multivariate regression to examine the associations between toxicity and users’ behaviors and to compare the behaviors of victims and other random users while considering the confounding factors. 
For numeric dependent variables we used Poisson multivariate regression models to test our hypotheses and for binary dependent variables, we employed logistic regressions. 
In all the models, we added all independent and control variables.

\textbf{H1: Toxicity victims are more likely to deactivate their accounts compared to random authors}. 
Victims might be so negatively affected that they decide to deactivate their accounts. 
Even though one toxicity attack might not trigger such behavior, we examine 
whether users tend to deactivate their social media accounts after receiving toxic comments on their posts. 
We only found 51 toxicity victims had deactivated their accounts, only contributing to 1\% of all CTV in our dataset. While this seems like a rare phenomenon, considering such drastic action is still alarming. 
We ran a logistic regression model having \emph{deactivated} as the dependent variable.
\emph{The results presented in Table~\ref{regression} (column H1) suggested that there is no statistical difference in account deactivation among toxicity victims and other users, and there is no evidence to support H1.} 

\textbf{H2: Toxicity victims are more likely to switch their accounts to private mode}. 
We examined switching to private mode as a \emph{countermeasure} activity because in this mode, only the user's followers can reply under their tweets. 
Twitter recently added \emph{circles} as a mechanism that allows  limiting audience using more fine-grained policies, however, this option was not available during our data collection. 
We found that switching accounts to the private mode occurred in only 0.6\% of CTV, and the logistic regression analysis presented in Table~\ref{regression} (column H2) revealed that there is no association between becoming \emph{private} and any type of \emph{toxicity}, rejecting H2. This can have various reasons, e.g., users might go private for other reasons, like when they post about something sensitive or private. 

\textbf{H3: Toxicity victims are more likely to engage in
conversations.} 
Prior study~\cite{ericsti2019reactions} showed that victims might \emph{negotiate} by communicating with the instigator to stop the aggressive behavior. H3 aims to test whether toxicity victims tend to \emph{negotiate} their points of view by engaging in conversations.
We measured engagement using \emph{\#root\_author\_replies} as the dependent variable. 
The results of the Poisson regression model are shown in Table~\ref{regression} (column H3). They indicate that there is a positive significant association between the \emph{nested\_toxicity} and the \emph{\#root\_author\_replies} ($p < 0.001$), as well as between the \emph{toxicity\_to\_others} and the \emph{\#root\_author\_replies} ($p < 0.001$). However, the relationship between \emph{direct\_toxicity} and \emph{\#root\_author\_replies} is negatively significant ($p < 0.001$). In other words, toxicity victims are more likely to engage in the conversation with more toxic nested and toxic replies to others, compared to random authors. This might indicate that root authors engage more in cases when discussion develops within the conversation, while they do not tend to respond to direct toxicity. 

Interestingly, \emph{verified} accounts tend to engage less in conversations compared to other users ($p < 0.001$), which can be due to the great number of replies they receive on their posts, or that they believe \emph{any attention is good attention}~\cite{palomino2020any,geiss2017any}. 
Moreover, young and more identifiable accounts who provide location and URL ($p < 0.001$), and have larger description length tend to engage more in toxic conversations compared to other users, which might indicate that more \emph{identifiable} users care more about their account reputability. Finally, there is a significant negative correlation between the \emph{\#root\_author\_replies} in the conversation and \emph{root\_toxicity} ($p < 0.001$). This shows that victims who started the conversation with a toxic tweet engage
less in conversations compared to victims whose conversations did not start with the toxic tweet.
\emph{In conclusion, this analysis supports H3 as it shows that toxicity victims are more likely to engage in the conversation when receiving more toxicity.} 

\textbf{H4: Toxicity victims are more likely to engage in
conversations if receiving toxic replies from a larger number of toxicity instigators}.
This hypothesis examines whether victims tend to engage more in CTV if toxicity they receive is coming from a larger percentage of users involved in the conversation. 
Users might not react the same way if receiving one or many toxic replies from a single user vs many users.  
To test this hypothesis, we defined our independent variable, \emph{toxicity\_instigators}, as the percentage of unique users who are posting toxic replies to the total number of unique users in the conversation. 
We avoided using this variable together with our independent variables in other models due to multicollinearity, as number of toxicity instigators is correlated with the number of toxic replies in the conversation. The results of employing the Poisson regression model suggest that users are more likely to engage in the conversation if toxicity is coming from a larger number of toxicity instigators compared to other users ($p < 0.001$), and therefore H4 is supported. 

\textbf{H5: Toxicity victims are more likely to respond back in a toxic way}.
Some victims might try to get revenge by responding to toxic comments in a toxic way. 
The selected dependent variable of the Poisson regression model is \emph{\#root\_author\_toxic\_replies} in the conversation, while predictors are the same as in the previous models.  
The results of regression analysis, illustrated in Table~\ref{regression} (column H5), show that there is a positive significant correlation between the number of root authors' toxic replies and all three \emph{toxicity} variables ($p < 0.001$). That is, the more toxic replies posted by other users are involved in the conversation, toxicity victims are more likely to respond back in a toxic way compared to random root authors, and therefore H5 is supported. 
Furthermore, root authors are more likely to engage in a conversation in a toxic way if the conversation root is toxic compared to other users ($p < 0.001$). Similarly as in H3, we found that young ($p < 0.001$) and identifiable accounts which provide location ($p < 0.001$) and URL ($p < 0.01$) are more likely to respond in a toxic way. 
This can be explained as users whose identities are known to their social network might argue back about their point of view instead of letting others openly attack them and degrade their reputation.  

\textbf{H6: Toxicity victims are more likely to respond back in a toxic way if receiving toxic replies from a larger number of toxicity instigators}.
Toxicity victims might react differently depending on the number of unique toxicity instigators. 
We used \emph{toxicity\_instigators} as the independent variable and ran the Poisson regression model.
\emph{The results suggest that users are more likely to respond back in a toxic way if toxicity is coming from more toxicity instigators compared to other users ($p < 0.001$), supporting H6.}

\textbf{H7: Toxicity victims are more likely to delete their original posts compared to random authors}. 
Receiving toxic comments on a tweet, the users might regret sharing their thoughts and decide to delete their post (\emph{signs of getting silenced}). 
We ran logistic regression model having \emph{root\_tweet\_deleted} as a binary dependent variable.
\emph{As it is shown in Table~\ref{regression} (column H7), the test did not yield any significant evidence that root tweet is more likely to get deleted in CTV compared to COA, rejecting H7.}  
This can have some explanation as users might decide to delete their tweets for other reasons than receiving toxic replies, e.g., when the tweet is not relevant anymore, or for privacy and cleaning the traces of activities from the social media~\cite{mondal2016forgetting, turner2017like}.  
However, we found that root tweets that are toxic itself, showed by the control variable \emph{root\_toxicity}, are more likely to be deleted compared to non-toxic root tweets ($p<0.001$). 
It could be that authors of toxic roots decide to remove their posts because they realize that the content was not appropriate, or they might not receive support from other users as they expected. 

\textbf{H8: Replies in conversations with toxic replies are more likely to be deleted compared to conversations without toxic replies}. 
This hypothesis examines whether replies that belonged to the root author are being removed more in CTV compared to COA. 
We ran a Poisson model with \emph{\#deleted\_replies} by the root author as the dependent variable in the model. 
The results obtained from the regression analysis (presented in Table~\ref{regression} in column H8) suggest that there is a statistically significant positive relationship between \emph{\#deleted\_replies} that belonged to the root author and \emph{nested\_toxicity} ($p < 0.001$), supporting H8 partially. This indicates that the CTV with the higher percentage of toxic nested replies are more likely to experience a higher number of deleted comments that belonged to the root author compared to COA. 

\textbf{H9: Toxicity victims are more likely to unfollow toxicity instigators compared to random authors}. 
Unfollowing can be an example of employing a \emph{countermeasure}, as it can reduce the probability of receiving additional toxic comments on future posts. 
To further test whether root authors tend to unfollow the accounts who post toxic comments on their posts that are directed to them, we averaged the toxicity scores of all comments belonging to each replier within a conversation. 
Afterwards, we ran a logistic regression model where the binary dependent variable indicates whether the root author unfollowed the replier while independent variable is the repliers' average toxicity scores. 
\textbf{Additional control variables:} We used two sets of control variables in this model, including a set of control variables related to the root author and control variables describing the replier. Victims might perceive toxic replies differently depending on who the replies are. For example, victims might be more likely to unfollow instigators who are not identifiable accounts. 
\emph{The results in Table~\ref{testing-h9} showed that there is a statistically significant positive correlation between unfollowing the replier and replier's average toxicity score within a conversation ($p < 0.001$), supporting H9. }
Furthermore, significance of control variables suggest that root authors are more likely to unfollow \emph{identifiable} repliers with the lower number of followers ($p < 0.05$). 
\begin{table*}[!htbp] \centering 
  \caption{Results of Hypotheses Testing} 
  \label{regression} 
  \resizebox{0.8\textwidth}{!}{%
\begin{tabular}{lllllll} 
\\[-1.8ex]\hline 
\hline \\[-1.8ex] 
 & \multicolumn{6}{c}{\textit{Dependent variable:}} \\ 
\cline{2-7} 
\\[-1.8ex] & Deactivated & Private & Author Replies & Author Toxic Replies & Root Deleted & Deleted Comments \\ 
\\[-1.8ex] & H1 (\textit{logistic}) & H2 (\textit{logistic}) & H3 (\textit{Poisson}) & H5 (\textit{Poisson}) &  H7 (\textit{logistic}) & H8 (\textit{Poisson}) \\ 
\hline \\[-1.8ex] 
 Direct Toxicity & 0.14 (0.23) & $-$0.17 (0.2) & $-$0.21$^{***}$  (0.02) & 0.45$^{***}$ (0.05) & 0.07 (0.11) & $-$0.38 (0.57)\\ 
  Nested Toxicity & 0.85 (0.81) & 1.06 (0.64) & 1.5$^{***}$ (0.04) & 3.01$^{***}$ (0.09)  & 0.62 (0.84) & 3.31$^{***}$  (0.74) \\ 
  Toxicity to Others & 0.07 (0.67) & 0.81 (0.42) & 0.69$^{***}$ (0.02) & 0.86$^{***}$ (0.09) & $-$0.81 (0.99) & 0.28 (1.11) \\ 
Width & 0.02 (0.03) & $-$0.02 (0.04) & $-$0.002$^{***}$ (0.0) & $-$0.001 (0.0) & $-$0.04$^{***}$ (0.01)  & 0.01 (0.04) \\ 
  Depth & $-$0.06 (0.06) & $-$0.16$^{**}$ (0.06)  & 0.03$^{***}$ (0.0) & 0.03$^{***}$ (0.001) & $-$1.36$^{***}$ (0.06)  & 0.05 (0.05)\\ 
Size & $-$0.02 (0.03) & $-$0.01 (0.03) & 0.004$^{***}$ (0.0) & 0.003$^{***}$ (0.0) & 0.04$^{***}$ (0.01) & $-$0.01 (0.04) \\ 
  Root Toxicity & 0.31 (0.16)  & 0.17 (0.14) & $-$0.05$^{***}$ (0.01) & 0.82$^{***}$ (0.03) & 0.4$^{***}$ (0.07) & $-$0.76 (0.52) \\ 
  Followers & 0.0 (0.0) & 0.0 (0.0)  & 0.0$^{***}$ (0.0) & $-$0.0$^{**}$ (0.0) & 0.0$^{*}$ (0.0) & 0.0 (0.0)  \\ 
  Friends & $-$0.0 (0.0) & $-$0.0 (0.0) & 0.0$^{**}$ (0.0) & $-$0.0 (0.0) & 0.0$^{***}$ (0.0) & 0.0 (0.0) \\ 
  Num Tweets & 0.0 (0.0) & 0.0$^{***}$  (0.0) & 0.0$^{***}$ (0.0) & 0.0$^{***}$ (0.0) & $-$0.0$^{***}$ (0.0) & $-$0.0 (0.0) \\ 
  Listed Counts & $-$0.0 (0.001) & $-$0.003$^{*}$ (0.002) & $-$0.0$^{***}$ (0.0) & $-$0.0$^{*}$ (0.0) & $-$0.0 (0.0) & $-$0.05$^{*}$ (0.02) \\ 
  Description Length & $-$0.01$^{***}$ (0.001) & $-$0.01$^{***}$ (0.001) & 0.001$^{***}$ (0.0) & 0.0 (0.0) & $-$0.01$^{***}$ (0.001) & $-$0.01 (0.002) \\ 
  Verified & $-$0.11 (0.76) & $-$1.34 (1.02) & $-$0.59$^{***}$ (0.03) & $-$1.03$^{***}$ (0.17)& 0.08 (0.17) & $-$376.87 (393.53) \\ 
  Account Age & $-$0.15$^{***}$ (0.02) & $-$0.11$^{***}$ (0.01) & $-$0.03$^{***}$ (0.001) & $-$0.05$^{***}$ (0.004) & $-$0.06$^{***}$ (0.01) & $-$0.11$^{**}$ (0.04) \\ 
  Location & $-$0.12 (0.12) & 0.36$^{***}$ (0.11) & 0.16$^{***}$ (0.01) & 0.18$^{***}$ (0.03) & $-$0.02 (0.06) & 0.2 (0.26) \\ 
  URL & 0.11 (0.12) & 0.02 (0.09) & 0.04$^{***}$ (0.01) & 0.07$^{**}$ (0.03) & 0.25$^{***}$ (0.05) & $-$0.18 (0.25) \\ 
 \hline \\[-1.8ex] 
Observations & 73,728 & 73,728 & 79,799 & 79,799 & 73,728 & 73,728 \\ 
Log Likelihood & $-$2,037.486 & $-$3,109.605 & $-$139,221.100 & $-$20,675.760 & $-$7,238.627 & $-$613.235 \\ 
\hline 
\hline \\[-1.8ex] 
\textit{Note:}  & \multicolumn{6}{r}{$^{*}$p$<$0.05; $^{**}$p$<$0.01; $^{***}$p$<$0.001} \\ 
\end{tabular} 
}
\end{table*} 
\section{Discussions and Limitations}
This study presents interesting findings that demonstrate the association between toxicity and users' behavioral responses. Prior work has shown that toxicity has negative effects on users' mentality and well-being, and with this work, we showed that toxicity affects victims' online behaviors. Even though approximately one third of the victims just ignored toxicity on their posts, which is also considered as a (lack of) reaction, others performed one or more reactions that are not necessarily considered as positive. For example, we found that victims are more likely to respond back in a toxic way. 
We also found that he location of toxic replies plays an important role when it comes to behavioral reactions of the victims. For example, victims tend to not remove their initial posts and conversation comments if receiving toxic \emph{direct} comments, while they are more likely to remove them receiving toxic \emph{nested} replies. 
Also, this example shows that some victims tend to \emph{avoid} further argument by deleting their comments, which indicates that toxicity has an impact on users' behaviors.
Furthermore, we showed that similar victims can be clustered into groups based on their behavioral reactions and we also observed that some reactions tent to appear together. For example, we noticed that some victims unfollowed toxicity instigators and engaged in the conversation in a toxic way, while victims in the other cluster deleted the root tweet and unfollowed instigators. This implies that some users might share some psychological traits triggering the similar reactions of these users, which can be investigated in a future work. 

\textbf{Limitations:} We solely relied on the Perspective API model to identify toxic tweets, and our data thus inherits its shortcomings or biases. 
Additionally, we only considered tweets in our dataset which were written in English, and thus our analysis does not account for toxic behaviour in other languages. Studying the timelines of the accounts and establish a variable of the quantity of toxicity received could help overall understanding of social media users behaviors. 

\textbf{Broader impacts:} The primary aims of moderation and intervention techniques are to detect and remove toxic content and decrease the impact of toxic behavior online
~\cite{finkelhor2000online,ericsti2019reactions}.   
We argue that our study can help with both aims. First, toxicity can be defined by context and culture; by analyzing the content that evokes victims' reactions, social media platforms might be able to implement more effective detection and moderation methods. 
We also argue that such analysis can help social media platforms to implement a targeted intervention method, since it can help identify vulnerable individuals, who might need more support from the community or the platform. 

\section{Conclusion}
One of the main contributions of this study is the analysis of victims' behavioral reactions against toxic replies on social media, in particular Twitter, in terms of \emph{avoidance}, \emph{countermeasures}, \emph{negotiation} and \emph{revenge}.
In conclusion, our results contribute to the general understanding of the impact of toxic replies on social media users' online behaviors. This study can help develop efficient intervention mechanisms to help mitigate the negative consequences of receiving toxic replies on social media. 



\bibliographystyle{ACM-Reference-Format}
\bibliography{references}

\appendix
\section{Longitudinal Analysis of Account Deactivation, Switching to Private Mode, and Deleting Root Tweets} 
\label{appendixA}
\begin{figure}[h!]
     \centering
     \begin{subfigure}[b]{0.3\textwidth}
         \includegraphics[width=\textwidth]{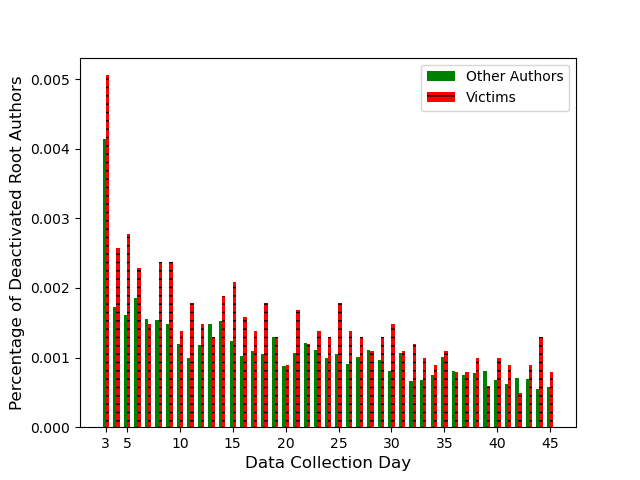}
         \caption{Percentage of Deactivated Accounts}
         \label{daily_deact}
     \end{subfigure}
     \hfill 
     \begin{subfigure}[b]{0.3\textwidth}
         \includegraphics[width=\textwidth]{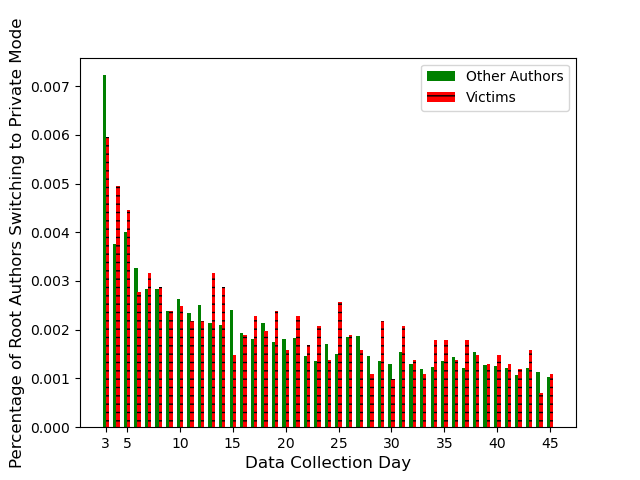}
         \caption{Percentage of accounts that went private}
         \label{daily_private}
     \end{subfigure}
     \hfill 
     \begin{subfigure}[b]{0.3\textwidth}
         \includegraphics[width=\textwidth]{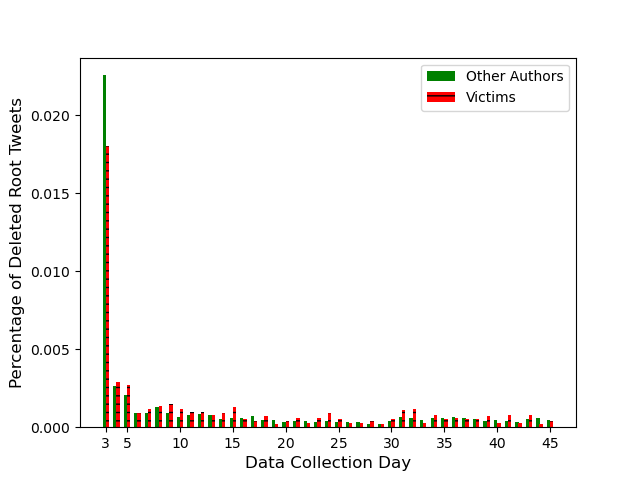}
         \caption{Deleted replies per Day}
         \label{ddaily_deletion}
     \end{subfigure}
        \caption{Daily Reactions among Toxicity Victims and other Users}
        \label{daily-reactions}
        
\end{figure}
Figure~\ref{daily-reactions} illustrated the distribution of daily percentages of newly deactivated accounts, private accounts, and deleted tweets for nine weeks. Interestingly, all figures demonstrate spikes of these three reactions on day three, for both victims and other authors, while the distribution on other days is mostly uniform. Even though the results show that there is no significant evidence that toxicity impacts account deactivation, the percent of victims who deactivated their accounts is around 1\%, which is higher compared to other authors who deactivated their accounts (0.4\%) on day three (as illustrated in Figure~\ref{daily_deact}). In case of switching the accounts to the private mode, Figure~\ref{daily_private} shows that, on day three, a higher percentage of other authors went private compared to victims (0.6\% vs. 0.8\%). There can be many reasons behind such occurrence, e.g., some users share a new posts and make their profiles public in order to receive a greater attention to that specific post. After couple of days, they might go back to being private. In terms of root tweet deletion shown in Figure~\ref{ddaily_deletion}, more root tweets belonging to COA are getting deleted on day three compared to roots of CTV. The reason behind this might be that deletion of the root tweet does not really remove the conversation comments, so there is really nothing that victims can do it terms of toxic comments to their root. It is also interesting that root tweets that initiated COA get deleted at a higher rate. It might happen due to authors not getting desired amount of feedback, i.e., comments, likes or retweets, or for privacy reasons they decide to remove their activities from the social media.

\section{The results for H9} 
H9: Toxicity victims are more likely to unfollow toxicity instigators compared to random authors. 
\begin{table}[!htbp] \centering 
  \caption{Results of Testing H9} 
  \label{testing-h9} 
  \resizebox{0.75\columnwidth}{!}{%
\begin{tabular}{@{\extracolsep{5pt}}lc} 
\\[-1.8ex]\hline 
\hline \\[-1.8ex] 
 & \multicolumn{1}{c}{\textit{Dependent variable:}} \\ 
\cline{2-2} 
\\[-1.8ex] & Unfollowed \\ 
 & Logistic Regression \\ 
\hline \\[-1.8ex] 
 Score & 0.405$^{***}$ (0.044) \\ 
  Replier Followers & $-$0.0$^{*}$ (0.0) \\ 
  Replier Friend & 0.0$^{**}$ (0.0)\\ 
  Replier Num Tweets & 0.0$^{***}$ (0.0)\\ 
  Replier Listed Counts & 0.0001 (0.0001)\\ 
  Replier Description Length & $-$0.001$^{***}$ (0.0002)\\ 
  Replier Verified & $-$0.082 (0.113)\\ 
  Replier Account Age & $-$0.051$^{***}$ (0.002) \\ 
  Replier Location & 0.411$^{***}$ (0.018) \\ 
  Replier URL & 0.341$^{***}$  (0.016) \\ 
  Author Followers & $-$0.0$^{***}$ (0.0) \\ 
  Author Friends & 0.0$^{***}$ (0.0) \\ 
  Author Num Tweets & 0.0$^{***}$ (0.0)\\ 
  Author Listed Count & $-$0.0004$^{***}$ (0.0001)\\ 
  Author Description Length & $-$0.005$^{***}$ (0.0002) \\ 
  Author Verified & $-$0.529$^{***}$ (0.071) \\ 
  Author Age & $-$0.072$^{***}$ (0.002) \\ 
  Author Location & 0.306$^{***}$ (0.018) \\ 
  Author URL & $-$0.167$^{***}$ (0.016)\\ 
 \hline \\[-1.8ex] 
Observations & 266,937 \\ 
Log Likelihood & $-$65,836.860 \\ 
\hline 
\hline \\[-1.8ex] 
\textit{Note:}  & \multicolumn{1}{r}{$^{*}$p$<$0.05; $^{**}$p$<$0.01; $^{***}$p$<$0.001} \\ 
\end{tabular} 
}
\end{table} 

\end{document}